\documentclass[12pt]{iopart}
\usepackage{graphicx}
\usepackage{dcolumn}
\usepackage{bm}
\usepackage{acronym}
\usepackage[colorlinks,citecolor=blue,urlcolor=blue,bookmarks=false,hypertexnames=true]{hyperref}
\usepackage{aas_macros}

\acrodef{GW}{gravitational wave}
\acrodef{PE}{Parameter Estimation}
\acrodef{MCMC}{Markov Chain Monte Carlo}
\acrodef{DOF}{degrees of freedom}
\acrodef{PSD}{power spectral density}
\acrodef{SNR}{signal-to-noise ratio}

\newcommand{\D}{\mathrm{d}}

\begin{document}

\title[]{VAMANA: Modeling Binary Black Hole Population with Minimal Assumptions}

\author{Vaibhav Tiwari}

\address{Cardiff School of Physics and Astronomy, Cardiff University, Queens Buildings, The Parade, Cardiff CF24 3AA, UK.}
\ead{tiwariv@cardiff.ac.uk}

\begin{abstract}
The population analysis of compact binaries involves the reconstruction of some of the \ac{GW} signal parameters, such as, the mass and the spin distribution, that gave rise to the observed data. This article introduces VAMANA, which reconstructs the binary black hole population using a mixture model and facilitates excellent density measurement as informed by the data. VAMANA uses a mixture of weighted Gaussians to reconstruct the chirp mass distribution. We expect Gaussian mixtures to provide flexibility in modeling complex distributions and enable us in capturing details in the astrophysical chirp mass distribution. Each of the Gaussian in the mixture is combined with another Gaussian and a power-law to simultaneously model the spin component aligned with the orbital angular momentum and the mass ratio distribution, thus also wing us to capture their variation with the chirp mass. Additionally, we can also introduce broadband smoothing by restricting the Gaussian mixture to lie within a threshold distance of a predefined reference chirp mass distribution. Using simulated data we show the robustness of our method in reconstructing complex populations for a large number of observations. We also apply our method to the publicly available catalog of \ac{GW} observations made during LIGO's and Virgo's first and second observation runs and present the reconstructed mass, spin distribution, and the estimated merger rate of binary black holes.
\end{abstract}

%
%
%
%
%

\section{Introduction}
The tally on the number of observed binary black hole mergers is increasing \cite{2019PhRvX...9c1040A, 2019ApJ...872..195N, 2020ApJ...891..123N, 2019PhRvD.100b3007Z, 2020PhRvD.101h3030V, o3a_catalog, o3a_rnp}. Currently, data until the end of the first half of the third observation run is publicly available when there were around fifty confirmed observations \cite{losc}. Multiple observations are facilitating the modeling of population level properties of these binaries. Although the first results presenting the reconstructed mass, spin and redshift distribution show large credible intervals amounting up to two orders of magnitude, the uncertainties are expected to reduce as the catalog of observations grow in number \cite{o3a_rnp}. However, some of the factors contributing to this uncertainty may not be completely mitigated by only increasing the catalog size. This is primarily because the \ac{GW}s provide information on those signal parameters that directly impact the phase evolution. For most of the observations, the phase evolution of \ac{GW}s is dominated by the chirp mass ($\mathcal{M}$) and a combination of mass ratio and an effective-spin term ($\chi_{\mathrm{eff}}$). These are defined as
\begin{eqnarray}
\mathcal{M} &=& \frac{(m_1\,m_2)^{(3/5)}}{(m_1 + m_2)^{(1/5)}} \nonumber \\
\chi_{\mathrm{eff}} &=& \frac{s_{1z} + q\,s_{2z}}{1 + q},
\end{eqnarray}
where $q \equiv m_2/m_1$ is the mass ratio corresponding to the component masses of the binary $m_1$ and $m_2$, and $s_{1z}$ and $s_{2z}$ are the component of the spins aligned with the orbital angular momentum \cite{Cutler:1994ys, Poisson:1995ef, 1995PhRvL..74.3515B, 1996CQGra..13..575B, Damour:2001, baird-2013, 2016CQGra..33aLT01T}.
Similarly, the amplitude is directly affected by a combination of the redshift and the inclination angle \cite{2019ApJ...877...82U}. Moreover, chirp mass is completely degenerate with the redshift. Thus, there occurs a large uncertainty in the measurement of the component masses and the spins. 

Yet another source of uncertainty is introduced by the methodology used in reconstructing the population properties. If the methodology does not model the population properly, the reconstruction may miss the intricate features in the distribution. Multiple methods have been developed to model the population properties of the binary black holes \cite{2016PhRvX...6d1015A, 2019PhRvD.100d3012W, 2019ApJ...882L..24A, 2017PhRvD..95j3010K, 2019PhRvD.100d3012W, 2018ApJ...856..173T, 2019MNRAS.484.4008G, 2020arXiv200409700S, 2019arXiv191209708G, 2019MNRAS.484.4216R, 2018ApJ...863L..41F, 2020CQGra..37d5007K, 2020arXiv200500023K, 2020arXiv200807014R}. Often these methods assume a phenomenological shape for modeling the distribution of a property. For example, power-law distribution has been often used in modeling the component or the chirp mass distribution \cite{2016PhRvX...6d1015A, 2019PhRvD.100d3012W, 2019ApJ...882L..24A, 2017PhRvD..95j3010K, 2019ApJ...878L...1P}. An alternative approach is to perform a model-independent fitting like described in \cite{2017MNRAS.465.3254M, 2019PhRvD.100h4041B, 2019MNRAS.488.3810P} (\cite{2019MNRAS.488.3810P} discuss efficient modeling of the masses and spins on simulated data that does not suffer from selection bias).

In this article we describe VAMANA, a mixture model framework for reconstructing the population properties. We model the chirp mass distribution using a mixture of weighted Gaussians which we expect to be capable of modeling a variety of complex distributions \cite{meyer_1993}. Multiple binary formation channels can introduce features in the mass spectrum \cite{2020arXiv200500023K, 2020ApJ...893...35D, 2020arXiv200901861A, 2002ApJ...567..532H} and thus modeling the chirp mass distribution, which is not known apriori, using a mixture model makes intuitive sense. Additionally, we combine each of the Gaussian in the mixture with another Gaussian and a power-law to simultaneously model the spin component aligned with the orbital angular momentum and the mass ratio distribution. This also allows us to capture their variation with the chirp mass. We further discuss the method in section \ref{method} and present results for modeling performed on simulated and publicly available data in section \ref{results}. VAMANA is available online \cite{vamana_github}.

\section{Method}
\label{method}

The methodology to model population properties of merging compact binaries has been discussed in multiple publications \cite{2019MNRAS.486.1086M, 2018ApJ...868..140T, 2019PASA...36...10T}. Following \cite{2018ApJ...868..140T}, the posterior on model hyper-parameters is given by equation \ref{eq:pop_bayes},
\begin{equation}
p(\lambda | \{\bm{d}\}) \propto
\prod_{i=1}^{N_\mathrm{obs}} \frac{\int d\bm{\theta}\;p(\bm{d}_{i} | \bm{\theta} )\;p (\bm{\theta} | \lambda)}{\int \D\bm{\theta}\;p_{\mathrm{det}}(\bm{\theta})\; p(\bm{\theta}|\lambda)} \;p(\lambda)\, \equiv e^{\mathcal{L}} p(\lambda),
\label{eq:pop_bayes}
\end{equation}
where $\bm{d} \equiv \{ \bm{d_0}, \cdots, \bm{d}_{N_\mathrm{obs}}\}$ is the set of observations, $\lambda$ is the population model, $\bm{\theta}$ are the signal parameters and $p_\mathrm{det}(\bm{\theta})$ encodes the probability of an event with signal parameters $\bm{\theta}$ to be observed with confidence. $p(\lambda)$ is the prior probability on the model hyper-parameters and $\mathcal{L}$ is the log-likelihood. The analysis samples model hyper-parameters using \ac{MCMC} and thus does not require the normalisation constant for equation \ref{eq:pop_bayes}.
In practice equation \ref{eq:pop_bayes} is estimated using discrete samples. \ac{PE} analysis samples $p(\bm{d}_{i} | \bm{\theta} )$ for a population model $p (\bm{\theta} | \lambda_{\mathrm{PE}})$ \cite{2015PhRvD..91d2003V}, and large scale injection campaign are performed to estimate the sensitivity of the detector network for a population model $p(\bm{\theta} | \lambda_{\mathrm{inj}})$ \cite{2016PhRvX...6d1015A}. Both the numerator and the denominator are then calculated for a target population $p (\bm{\theta} | \lambda)$ using importance sampling \cite{2019MNRAS.486.1086M, 2018CQGra..35n5009T}. We note that the denominator in equation \ref{eq:pop_bayes}, $V(\lambda) \equiv \int \D\bm{\theta}\;p_{\mathrm{det}}(\bm{\theta})\; p(\bm{\theta}|\lambda)$, makes correction for the selection bias and is termed the sensitive volume for the population $p (\bm{\theta} | \lambda$) \cite{2018CQGra..35n5009T}.

\subsection{Choice of Signal Parameters}
The signal parameters are broadly categorised as a) intrinsic signal parameter: that are directly responsible for the orbital evolution of the binary, such as, masses, spins, tidal deformability, eccentricity, periastron distance, etc., and b) extrinsic signal parameters: that are observer-dependent, such as, luminosity distance, inclination of the binary from the line of sight, sky location, coalescence phase of the \ac{GW} signal and coalescence time of the \ac{GW} signal. Masses, spins, and redshift are signal parameters of primary interest for the binary black hole population. Usually, component masses, spin magnitude, and tilt-angle are chosen as the population property for reconstruction \cite{2019ApJ...882L..24A}. However, only a few parameters are directly responsible for the \ac{GW} signal's phase evolution. The dominant term for a binary's phase evolution is the chirp mass \cite{Cutler:1994ys}. The second leading order is a function of $\chi_{\mathrm{eff}}$ and $q$ \cite{Cutler:1994ys,Poisson:1995ef,baird-2013}. At this order, a high $\chi_{\mathrm{eff}}$ - low $q$ binary is indistinguishable from
a low $\chi_{\mathrm{eff}}$ - high $q$. The presence of this degeneracy between the masses and spins can be observed in most of the observations. Only for a small number of observations the individual spins or mass ratios are measurable. Taking the example of the recently announced asymmetric binaries GW190412 and GW190814 \cite{2020arXiv200408342T, 2020ApJ...896L..44A}, that were observed at a high signal-to-noise ratio and with evidence of contribution from the higher harmonics in the signal, the spins on the primary mass was measurable but limited constraints were put on either the secondary spin magnitude or the tilt-angle respectively. 

As not all the masses and spins are measured accurately, when reconstructing population properties, one may expect that by using priors close to the true astrophysical distributions the overall statistics will converge to the true astrophysical distribution for a large number of observations. Alternatively, it has been shown explicitly in \cite{2017PhRvL.119y1103V} and argued in \cite{2018ApJ...868..140T} that different priors on masses result in different inference on the spins and vice-versa. Moreover, the signal parameters are estimated with the assumption that the underlying noise in the instrument is Gaussian, however, as this is rarely the case, even with good priors the inference on the masses and spins may become biased and the reconstructed population may not converge to the true distribution even with a large number of observations. It is conceivable that a combination of improper prior with the non-convergence to the true distribution can result in biased inference on the population.

Most of the observations will not have signature for higher harmonics or precession \cite{2020PhRvD.102d1302F, 2020arXiv200704313M}. Thus, we choose signal parameters that are measured accurately. We choose chirp mass as a population property as it is measured accurately for a wide range of masses \cite{2015ApJ...804..114B, baird-2013}. Additionally, we choose mass ratio and component of the spins aligned with the orbital angular momentum as the other population properties. Furthermore, we assume the same distribution for both the spins. This choice abates the degeneracy in the masses and spins by favouring population models that can produce a small value of $\chi_{\mathrm{eff}}$ due to small value of $s_{1z}$ and $s_{2z}$ and disfavouring population models that can produce a small value of $\chi_{\mathrm{eff}}$ by a large positive value of $s_{1z}$ and a large negative value of $s_{2z}$ or vice-versa. Thus we choose to model the intrinsic signal parameters using distributions totaling three ($\theta \equiv {\mathcal{M}, q, s_z}$) \footnote{It has been suggested that the primary and the secondary black holes can have different spin distribution (e.g. see \cite{2018A&A...616A..28Q}); if future observations corroborate this suggestion, extending VAMANA to include $s_{2z}$ is straightforward.}.

\subsection{Modeling Using Gaussians}
Gaussian mixtures are often used in classifying or modeling the probability density of the observed data. A Gaussian mixture can also approximate a function with the hyper-parameters of the components calculated using expectation-maximisation \cite{10.2307/2030064, Reynolds2009}. In a Bayesian setting, likelihood is expressed as a sum of mixtures with an assumed prior distribution of the mixture hyper-parameters \cite{titterington_85}. The number of components in the mixture can be fixed with the number of components chosen based on the goodness of fit and the complexity of the mixture \cite{730550, doi:10.1177/0049124103262065} or can be flexible as informed by the data \cite{10.5555/3009657.3009736}.

In this analysis we model the population using a mixture of components. To effectively capture the variation of the mass ratio and the aligned spin component with the chirp mass, each component comprises of a Gaussian to model the the chirp mass, another Gaussian to model the aligned spin components and a power-law to model the mass ratio distribution. Equation \ref{eq:mixture} describes the distributions used in modeling the population. The notations used in this model are described in table \ref{table:params},

\begin{eqnarray}
p(\theta|\bm{\lambda}) \equiv \Pi(\mathcal{M}, q, s_{1z}, s_{2z}) = \nonumber\\\hspace{2cm}\sum_{i=1}^N w_i \;\phi(\mathcal{M}|\mu_i^{\mathcal{M}}, \sigma_i^{\mathcal{M}})\;\phi(s_{1z}, s_{2z}|\mu^{sz}_i, \sigma^{sz}_i)\;\mathcal{P}(q|\alpha_i^q, q_i^{min}, 1.0).
\label{eq:mixture}
\end{eqnarray}

\begin{center}
\begin{table}
\begin{tabular}{ | m{1cm} | m{7cm} | m{1cm} | m{7cm} |} 
\hline
$\bm{N}$ & Number of components in the mixture & $\mathbf{s_{2z}}$ & Second aligned spin component\\ 
\hline
$\mathbf{w_i}$\vspace{1mm} & Mixing weights\vspace{1mm} & $\bm{\mu^{sz}_i}$\vspace{1mm} & Mean of the Gaussians modeling the aligned spin distribution\\ 
\hline
$\bm{\phi}$ \vspace{1mm}& Normal distribution \vspace{1mm}& $\bm{\sigma^{sz}_i}$ \vspace{1mm} & Standard deviation on the Gaussian  modeling the aligned spin distribution\\ 
\hline
$\bm{\mathcal{M}}$& Chirp Mass & $\bm{\mathcal{P}}$ & Powerlaw distribution\\ 
\hline
$\bm{\mu_i^{\mathcal{M}}}$ \vspace{1mm}& Mean of the Gaussians modeling the chirp mass & $\bm{\alpha_i^q}$ \vspace{1mm}& Slope of the power-law \vspace{1mm}\\ 
\hline
$\bm{\sigma_i^{\mathcal{M}}}$\vspace{1mm} & Standard deviation of Gaussians modeling the chirp mass & $\bm{q_i^{min}}$\vspace{1mm} & Minimum value of the mass ratio (maximum is one)\\ 
\hline
$\bm{s_{1z}}$ & First aligned spin component & $\bm{\mathcal{L}}$ & Log likelihood\\
\hline
\end{tabular}
\caption{Description of notations used in describing the model.}
\label{table:hparams}
\end{table}
\end{center}
Probability distribution in equation \ref{eq:mixture} is extended to include the merger rate by incorporating the Poisson term \cite{extended_lkl}
\begin{equation}
p(\mu) = \frac{\mu^{-N_\mathrm{obs}}\mathrm{e}^{-\mu}}{N_\mathrm{obs}!},
\end{equation}
where $\mu = R\;V(\lambda)$ is the expected number of observations for the merger rate $R$.

The posterior on the hyper-parameters of the reconstructed population, $p (\bm{\theta} | \lambda$), are obtained by using Metropolis-Hastings sampling \cite{10.1093/biomet/57.1.97}. Proposals of hyper-parameters are made and acceptance probability is calculated using proposal distribution and the full joint density $\mathcal{L}$. 

\subsection{Constraints and Smoothing}

The proposed chirp mass in VAMANA is a combination of Gaussians with no constraints on their scale or location and thus to contain the error encountered in importance sampling while evaluating equation \ref{eq:pop_bayes} we constrain the Gaussians modeling the chirp mass distribution to have $\sigma_i^{\mathcal{M}}$ always larger than 0.05 $\mu_i^{\mathcal{M}}$ and the Gaussians modeling the aligned-spin distribution to have $\bm{\sigma^{sz}_i}$ larger than 0.05. This condition ensures that Gaussians have scales larger than the usual standard deviation for the parameter estimates of $\mathcal{M}$/$\chi_{\mathrm{eff}}$ for most of the observations.

To help with better observing features and trends we can also introduce a broad-band smoothing by using a reference population. We can perform this by iteratively changing the hyper-parameters of a simple phenomenological model that uses,
\begin{itemize}
\item a power-law with fixed cutoff range to model the chirp mass distribution, $p(\mathcal{M}) = \mathcal{P}(\mathcal{M}|\mathcal{M}_{\mathrm{min}}, \mathcal{M}_{\mathrm{max}}, \alpha^\mathcal{M}_{\mathrm{reference}})$, where we set the cut-offs at the first percentile of the chirp mass estimates of the lightest binary black hole observation ($\mathcal{M}_{s}^{\mathrm{min}}$) and at the eightieth percentile of the chirp mass estimates of the heaviest binary black hole observation ($\mathcal{M}_{s}^{\mathrm{max}}$). Although seemingly arbitrary, these are broad choices and have a negligible impact on the results as the \ac{PE} samples have negligible support outside this range. 
\item A truncated Gaussian with boundaries at $s_z^{\mathrm{max}} = \pm 0.99$ to model the spin distribution, $p(s_z) = \psi(s_z|\mu^{sz}_{\mathrm{reference}}, \sigma^{sz}_{\mathrm{reference}})$.
\item And a power-law distribution with boundaries at 0.1 and 1 to model the mass ratio distribution, $p(q) = \mathcal{P}(q|0.1, 1.0, \alpha^q_{\mathrm{reference}})$.
\end{itemize}
and identify the maximum likelihood fit as the reference population. There are four hyper-parameters in our phenomenological model. We update the values of the hyper-parameters to the ones drawn using normal distributions around the current values every-time the likelihood increases.

In equation \ref{eq:reff} we define a distance measure $r_{\mathrm{eff}}$, inspired from the idea of importance sampling,
\begin{equation}
r_{\mathrm{eff}} = \frac{1}{n} \frac{\sum_{i = 1}^{i = n} w_i}{\,\mathrm{max}(\bf{w})},\;\;\;\; w_i = \frac{\mathrm{P}(\mathcal{M}_i|\lambda_\mathrm{proposed})}{ \mathrm{P}(\mathcal{M}_i|\lambda_\mathrm{reference})},
\label{eq:reff}
\end{equation}
where $\mathbf{w} \equiv (w_0 \cdots w_n)$ are ratios of the probabilities calculated on $n$ chirp mass bins centered at $\mathcal{M}_i$. 
This measure is closely related to the Euclidean distance-squared between the reference population's chirp mass distribution and a proposed chirp mass distribution as defined in equation \ref{eq:measure} with $p = 2$ \cite{ESS},
\begin{equation}
L_p = \int \left(p(\mathcal{M}|\lambda_\mathrm{proposed}) - p(\mathcal{M}|\lambda_\mathrm{reference})\right)^p\;\D \mathcal{M}.
\label{eq:measure}
\end{equation}
We can expect the reference chirp mass distribution to have an $L_1$ value close to zero for the true chirp mass distribution for a large number of observations. Additionally, we don't expect our simple phenomenological model to fit the data very well, and thus $L_2$ value will be non-vanishing and will depend on the complexity of the true distribution. Hence we can employ the Gaussian mixture to explore the chirp mass distribution in the vicinity of the reference distribution and putting a threshold on the distance between the reference and the proposed chirp mass distribution provides broadband smoothing. Unlike a phenomenological function that gets modified throughout the chirp mass range, we explore all distributions -- expressible as the sum of weighted Gaussians --  that are within a distance measure from the reference chirp mass distribution. Although a threshold can be applied on either $r_{\mathrm{eff}}$ or $L_p$, we choose to put a threshold on $r_{\mathrm{eff}}$. $r_{\mathrm{eff}}$ lies between zero and one with a threshold of zero allowing the mixture of Gassusians complete freedom and a threshold of close to one requiring proposed chirp mass distribution to be close to the reference chirp mass distribution. 



\subsection{Priors and Proposals}

Table \ref{table:hparams} lists the prior applied on the hyper-parameters. Except for the location of Gaussians modeling the chirp mass and the merger rate, that follow a uniform-in-log prior, a uniform prior is applied on all the remaining hyper-parameters.  The range of all the priors is fixed except for the maximum value of $\bm{\sigma_i^{\mathcal{M}}}$ which is chosen proportional to $\bm{\mu_i^{\mathcal{M}}}/\sqrt{N}$ and for the maximum value of $\bm{\sigma^{sz}_i}$ which is chosen inversely proportional $\sqrt{N}$. With these choices, the mean of the chirp mass distribution corresponding to the hyper-parameter priors is approximately uniform-in-log and the priors on the chirp mass, aligned-spin, and mass ratio remain almost unchanged for a wide range of component number. This will also provide consistent scaling when the number of components is increased to model a bigger gravitational wave catalog in the future. The prior on the merger rate is scale-invariant and does not contribute to the posterior of other hyper-parameters\cite{uniforminlog}. The scale-invariant uniform-in-log prior on $\bm{\mu_i^{\mathcal{M}}}$ also keeps the prior intact in the event its ranges needs adjustment due to the addition of future observations with chirp masses outside the current range.

The scales of the Gaussians modeling the chirp mass and the aligned-spins are proposed using the $\chi^2$ distribution. To avoid Gaussians getting stuck at local maxima each proposal is made using a different value of the \ac{DOF} with these values drawn from a uniform distribution. A large value of \ac{DOF} proposes closer to the current value of the scale while a small value of \ac{DOF} proposes farther from the current value of the scale. The hyper-parameters $\bm{\alpha_i^q}$, $\bm{q_i^{min}}$, and merger rate are proposed by drawing from a normal distribution around the current values. The scale of the proposing Gaussians is pre-fixed. We use Dirichlet distribution to propose the mixing weights with DOF drawn from a uniform distribution.
\begin{center}
\begin{table}
\begin{tabular}{ | m{.8cm} | m{2.6cm} | m{2.1cm} | m{2.2cm} | m{2.1cm} | m{3.3cm} |} 
\hline
$\lambda$ & Prior & Minimum Value & Maximum Value & Proposal Distr.& Scale/DOF/Width of the Proposal Distr.\\ 
\hline
$\bm{\mu_i^{\mathcal{M}}}$ & Uniform in log & Fixed & Fixed & Uniform & See equation \ref{eq:prp_strategy}\\ 
\hline
$\bm{\sigma_i^{\mathcal{M}}}$ & Uniform & Fixed & $\propto \bm{\mu_i^{\mathcal{M}}}/\sqrt{N}$ & Chi-square & Variable\\ 
\hline
$\bm{\mu^{sz}_i}$ & Uniform & Fixed & Fixed & Uniform & See equation \ref{eq:prp_strategy}\\ 
\hline
$\bm{\sigma^{sz}_i}$ & Uniform & Fixed & $\propto 1/\sqrt{N}$ & Chi-square & Variable\\ 
\hline
$\bm{q_i^{min}}$ & Uniform & Fixed & Fixed & Normal & Fixed\\ 
\hline
$\bm{\alpha_i^q}$ & Uniform & Fixed & Fixed & Normal & Fixed\\ 
\hline
$\mathbf{w_i}$ & Uniform & Fixed & Fixed & Dirichlet & Variable\\ 
\hline
$R$ & Uniform in log & Fixed & Fixed & Normal & Fixed\\ 
\hline
\end{tabular}
\caption{Hyper-parameters used in VAMANA.}
\label{table:params}
\end{table}
\end{center}

\begin{figure}
\centering
\includegraphics[width=0.95\textwidth]{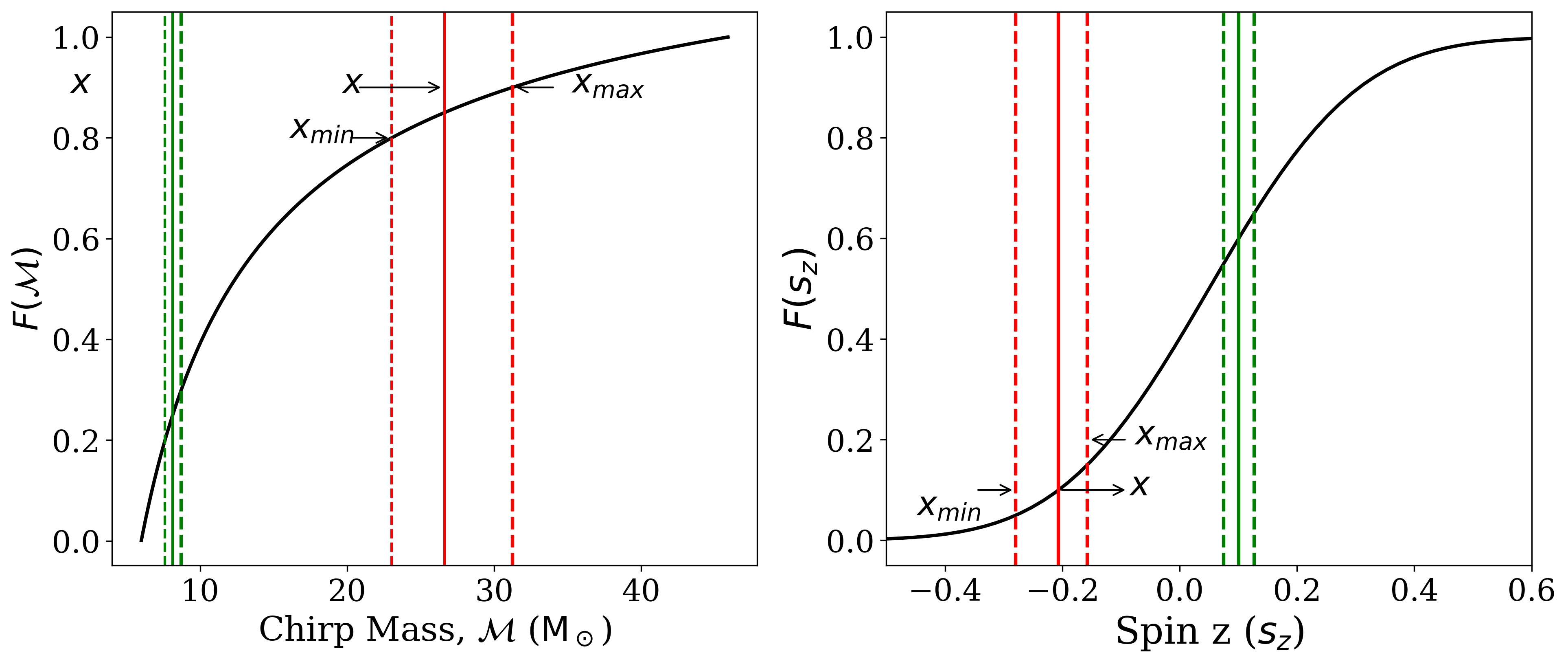}
\caption{An example elucidating the procedure to make proposals for $\bm{\mu_i^{\mathcal{M}}}$ and $\bm{\mu^{sz}_i}$. The locations of Gaussians modeling the chirp mass/aligned-spin are sampled by making proposals uniformly between $x_{\mathrm{min}}$ and $x_{\mathrm{max}}$. The procedure to calculate $x_{\mathrm{min}}$ and $x_{\mathrm{max}}$ is described in equation \ref{eq:prp_strategy}. The dashed lines in the plots show the interval in which the next proposal is made for a Gaussian located at the solid line. The width of the interval is smaller where the change in density is steep as shown by the green lines and larger where the change in density is shallow as shown by the red lines. The width of the support interval depend on the reference population as well as $\delta \mathcal{F}$.}
\label{fig:width_prp}
\end{figure}

Astrophysical chirp mass distribution is expected to have a fall-off similar to a power-law distribution, thus a shift in the location of a Gaussian at lower chirp mass will cause a larger change in the likelihood compared to the same shift in the location of a Gaussian located at a higher chirp mass value. The spins have been measured to be low and distributed normally \cite{o3a_catalog}, thus a shift in the location of a Gaussian at lower aligned/anti-aligned spin value will cause a larger change in likelihood compared to the same shift in the location of a Gaussian located at a higher aligned/anti-aligned spin value. To sample the posteriors efficiently the location of Gaussians are proposed using uniform distribution but with support range adjusted according to the current location. For the current location $x$, the support range [$x_\mathrm{min}, x_\mathrm{max}$] is calculated using the following prescription,
\begin{eqnarray}
x_{\mathrm{proposed}} &=& U(x_{\mathrm{min}}, x_{\mathrm{max}}), \nonumber \\
x_{\mathrm{min}} &=& \mathcal{F}^{-1}(\mathrm{maximum}(0, \mathcal{F}(x) - \delta \mathcal{F})), \nonumber \\
x_{\mathrm{max}} &=& \mathcal{F}^{-1}(\mathrm{minimum}(1, \mathcal{F}(x) + \delta \mathcal{F})),
\label{eq:prp_strategy}
\end{eqnarray}
where $U$ is the uniform distribution, $\mathcal{F}$ is the cumulative density function and $\mathcal{F}^{-1}$ is the inverse distribution function corresponding to $\mathcal{P}_{\mathrm{ref}}(\mathcal{M})$ or $\psi_{\mathrm{ref}}(s_z)$. $\delta \mathcal{F}$ determines $x_{\mathrm{min}}$ and $x_{\mathrm{max}}$, and directly impacts the interval [$x_\mathrm{min}, x_\mathrm{max}$]. We show this pictorially in figure \ref{fig:width_prp}.

\begin{figure}
\centering
\includegraphics[width=0.95\textwidth]{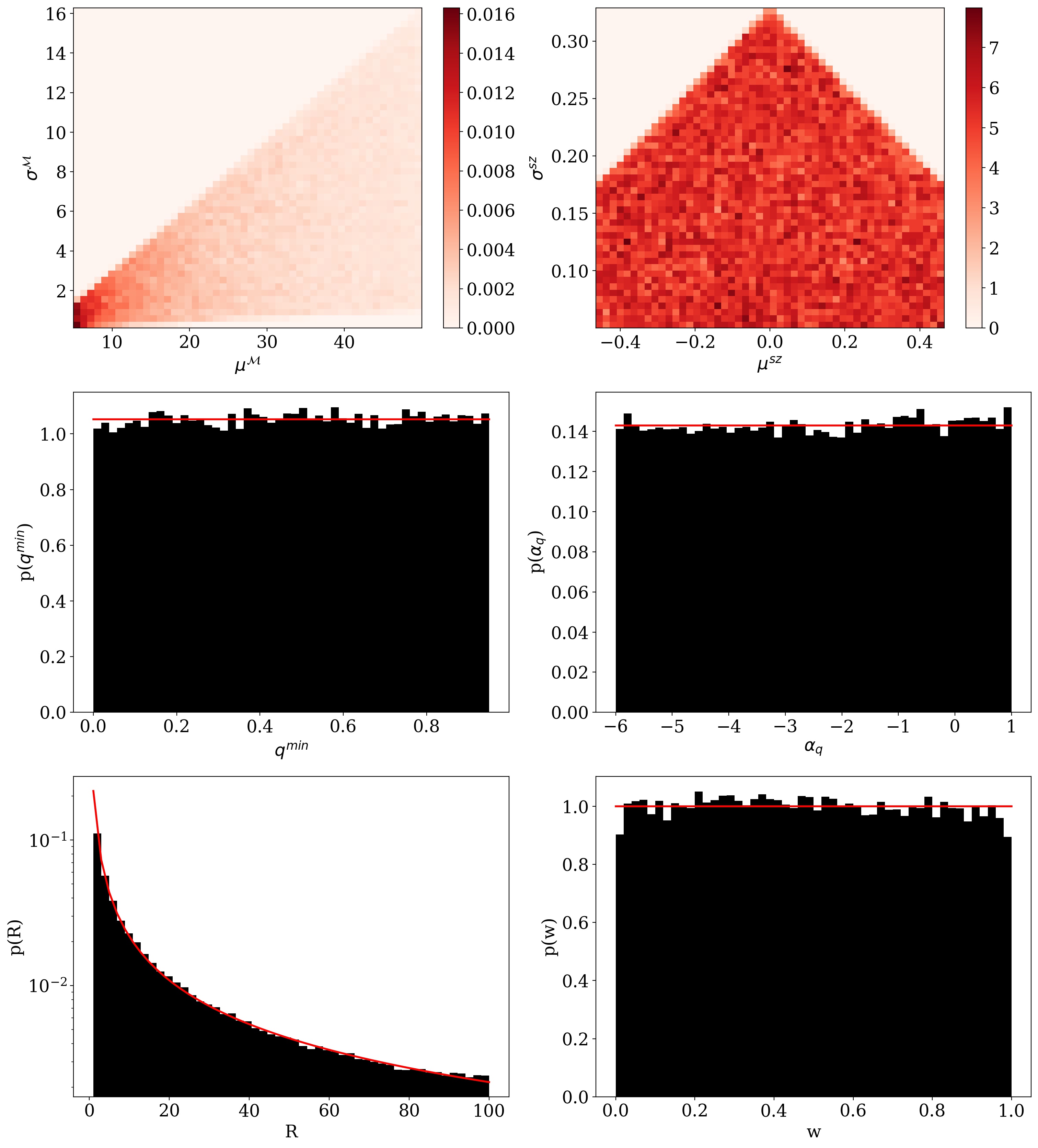}
\caption{An example prior distribution for the case of two Gaussians obtained by performing the analysis with a flat likelihood and no smoothing. The red curves are the expected distributions.}
\label{fig:prior}
\end{figure}

We have verified that the analysis reproduces the priors for the case of flat likelihood. We show this in figure \ref{fig:prior}. The number of Gaussians and the ranges or the hyper-parameters will need to be modified as catalog for the observations grows in size. We discuss this further in the context of the presented results in subsection \ref{sec:o1o2}.

\section{Results}
\label{results}
In this section, we discuss the robustness of VAMANA in reconstructing complex distributions using toy models that mimic the primary features of a full analysis. We also apply the analysis on the publicly available data and present the reconstructed mass and spin distribution, and the estimated merger rate for binary black holes.

\subsection{Toy Model}
\begin{figure}
\centering
\includegraphics[width=0.95\textwidth]{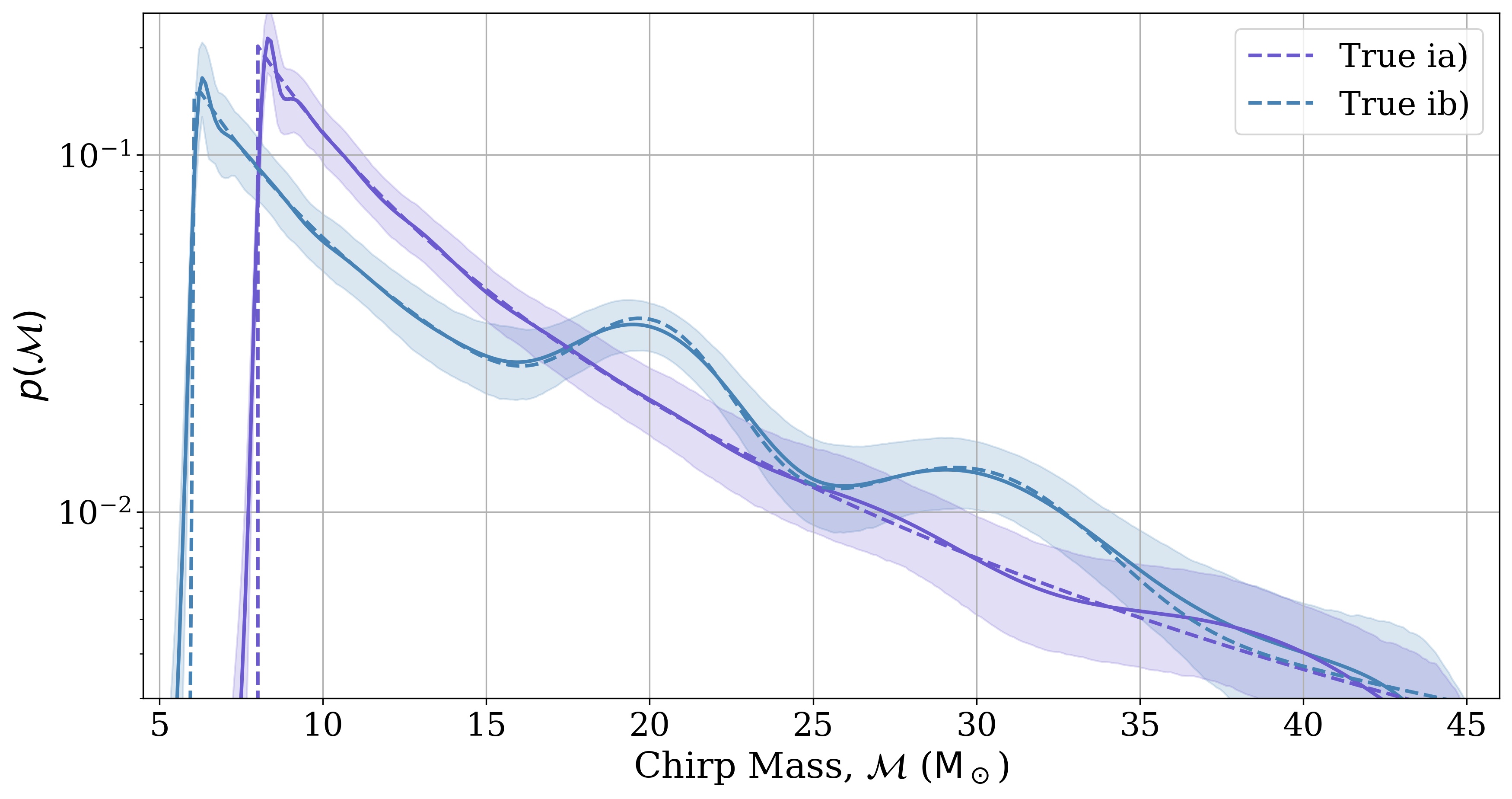}
\includegraphics[width=0.95\textwidth]{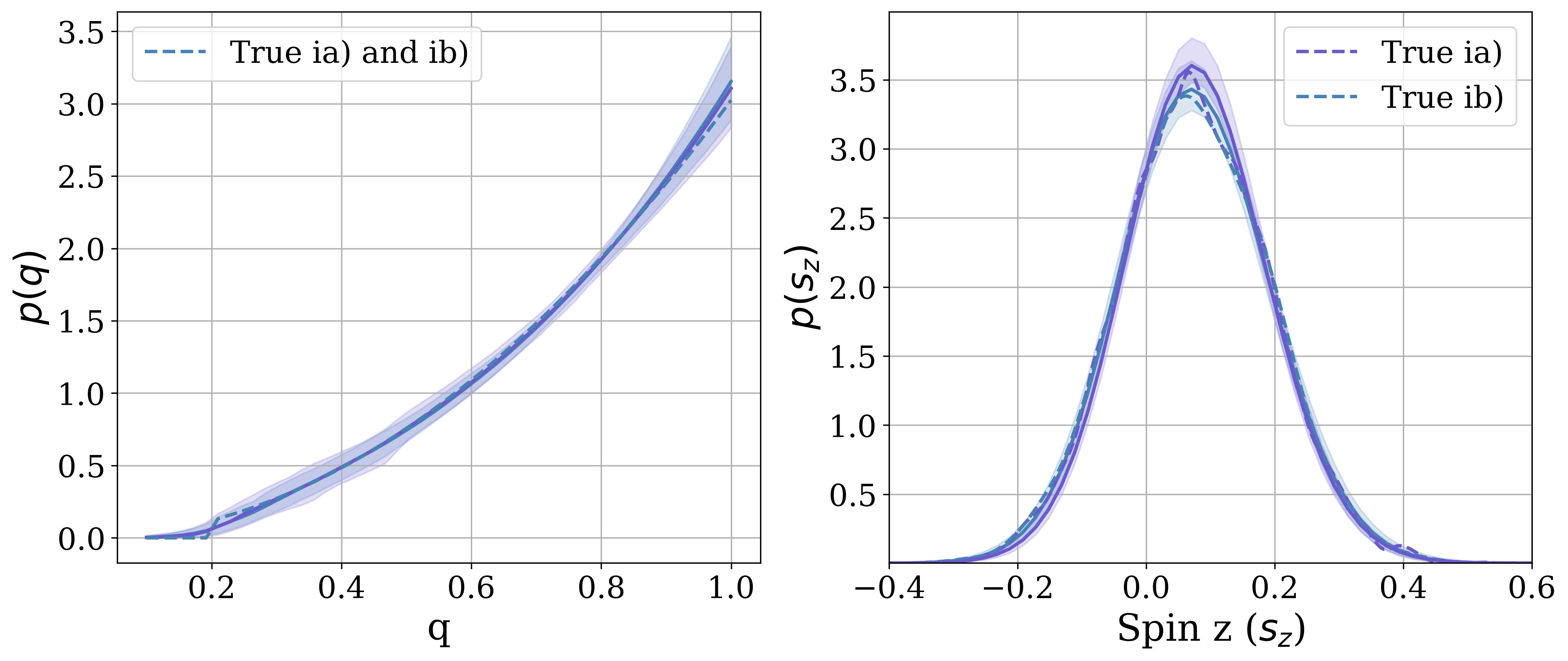}
\caption{The reconstructed top) chirp mass, bottom) mass ratio and aligned spin distribution for the toy models described in equations \ref{eq:toy_pop}. The curves are the mean distribution with shaded region representing the 90\% credible interval. The dashed curves are the true distributions. The reconstructed mass ratio and aligned spin distributions distribution have been marginalised over the chirp mass.}
\label{fig:mch_q_sz}
\end{figure}

\begin{figure}
\centering
\includegraphics[width=0.95\textwidth]{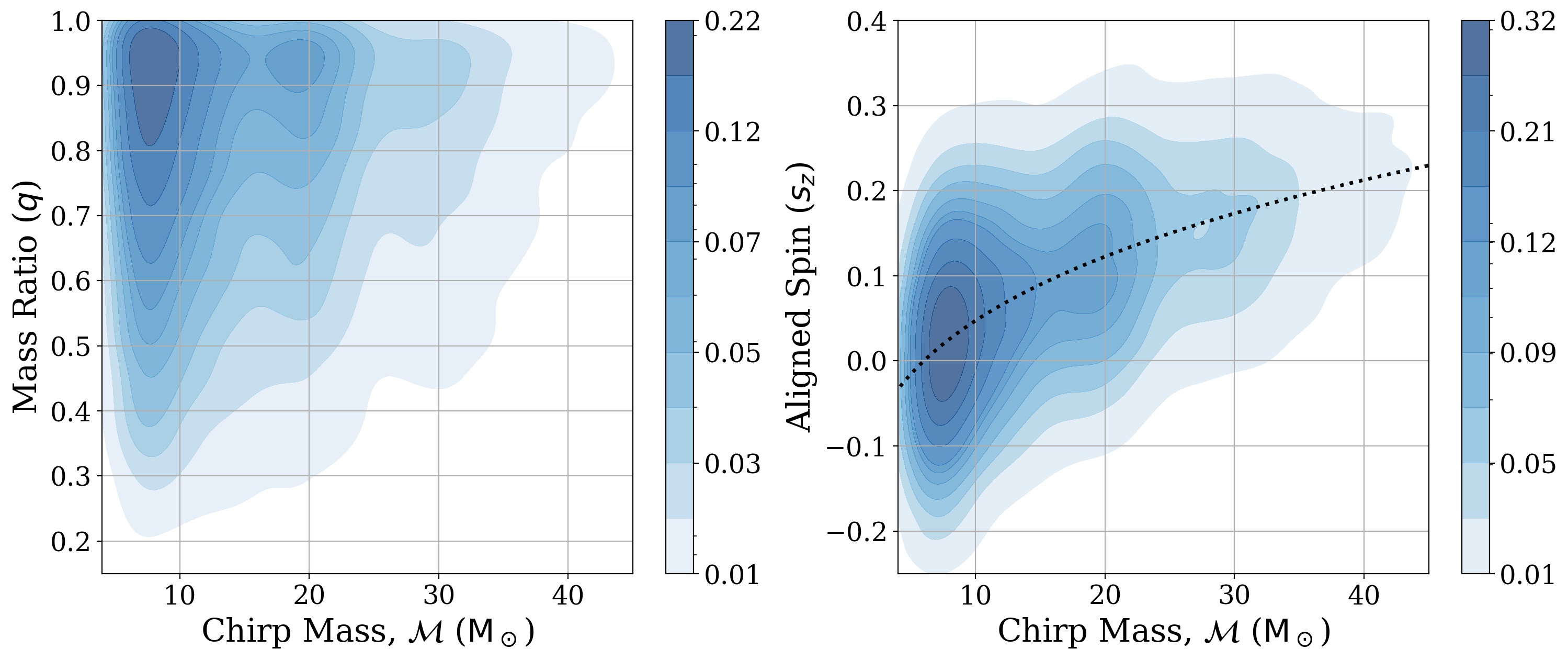}
\caption{The plot shows the reconstruction of aligned spin and mass ratio as dependent on the chirp mass for the model $ib)$. There are 10 contours in each plot showing equally spaced credible intervals with the first one for the 5\% confidence and the last one for the 95\% confidence.}
\label{fig:mch_sz}
\end{figure}

In an analysis with the real data, the presence of measurement uncertainty requires the likelihood to be marginalised over it and presence of selection effect requires proper re-scaling of the density. We have verified that both these procedures are performed accurately. Thus, we directly focus on the modeling capability of the analysis by concocting two complex toy model populations with distribution defined as,
\begin{eqnarray}
ia)\; p(\mathcal{M}) &=& \mathcal{P}\left(\mathcal{M}|\mathcal{M}_{\mathrm{min}} = 8.0M_\odot, \mathcal{M}_{\mathrm{max}} = 46.0M_\odot, \alpha = 2.5\right) \nonumber \\
&\mathrm{or}& \nonumber \\
ib)\;p(\mathcal{M}) &=& 0.85 \times\mathcal{P}\left(\mathcal{M}|\mathcal{M}_{\mathrm{min}} = 6.0M_\odot, \mathcal{M}_{\mathrm{max}} = 46.0M_\odot, \alpha = 2.0\right) \nonumber \\
&+& 0.1\times\phi\left(\mathcal{M}|\mu = 20M_\odot,\, \sigma = 2M_\odot\right) \nonumber \\ && +0.05\times\phi\left(\mathcal{M}|\mu = 30M_\odot,\, \sigma = 3M_\odot\right) \nonumber \\
ii)\;p(s_z) &=& \phi\left(s_z|\mu = \left(\left(\frac{\mathcal{M}}{5.0}\right)^{1/4} - 1\right)/3, \sigma = 0.1\right) \nonumber \\
iii)\;p(q) &=& \mathcal{P}\left(q|q_{\mathrm{min}} = 0.1, q_{\mathrm{max}} = 1.0, \alpha = -2.0\right).
\label{eq:toy_pop}
\end{eqnarray}
and simulating data directly from this population. For these analysis we generate 1000 simulated data points and use 9 Gaussians to reconstruct the population. The primary goal of the analysis is to verify a bias free methodology. Figure \ref{fig:mch_q_sz} plots the reconstructed population and figure \ref{fig:mch_sz} plots the reconstructed $\mathcal{M} - s_z$ and $\mathcal{M} - q$ distribution. The reconstructed distribution show excellent agreement with the true distribution.

We estimate the significance of any feature extracted by the analysis by making comparisons. For example, we can compare the significance of the two peaks in the reconstructed mean by comparing them with the underlying power-law distribution in the $ib)$ distribution. For this case we obtain a log Bayes factor of 98 and 43 respectively, i.e. the peaks are highly significant compared to the underlying power-law distribution. On the other hand, our $ia)$ distribution reconstructs a seemingly spurious peak centered at $38 M_\odot$. However, we calculate a log Bayes factor of around 1 when we compare this peak with the true distribution, i.e. this peak is barely worth mentioning.

\subsection{Reconstruction Using Observed Gravitational Waves}
\label{sec:o1o2}
In this section, we further discuss methodology by presenting the results obtained for the observations made during LIGO's and Virgo's first and second observation run \cite{2019PhRvX...9c1040A}. The analysis that uses all publicly available observations is discussed in a separate article \cite{2020arXiv201104502T}. We only select the events with a false alarm rate of at most once in five years in PyCBC or GstLAL search analysis \cite{pycbc, gstlal}. \ac{PE} samples of these observations are publicly available along with the prior used in producing these samples \cite{o1o2pe}. Parameter estimation analysis samples were generated using a stochastic sampler LALInference \cite{2015PhRvD..91d2003V}. Independent searches have reported few more \ac{GW} observations \cite{2019ApJ...872..195N, 2020ApJ...891..123N, 2019PhRvD.100b3007Z, 2020PhRvD.101h3030V} but we leave these observations out until a unified framework is in place that can consistently include observations made by many independent searches. Sensitive volume is estimated on the recovered injections that follow the power-law distribution in chirp mass and mass ratio, and uniform distribution in aligned spin components. Injections are distributed uniformly in extrinsic signal parameters, except the redshift, for which they are distributed uniform-in-comoving volume. The recovered injections are defined as the ones that cross a network \ac{SNR} of 9.0 on a given power spectral density (PSD). We use multiple PSDs chosen uniformly over the observation time. The threshold of 9.0 is chosen as all the observations have been observed at a higher \ac{SNR} by the search analysis, moreover, a simple quadrature sum suggests that the contribution from instrument noise is low at this \ac{SNR} threshold enabling our method in estimating \ac{SNR} similar to an actual search analysis. However, we expect the sensitive volume estimation to be approximate and may result in a slightly biased reconstruction.

\begin{figure}
\centering
\includegraphics[width=0.95\textwidth]{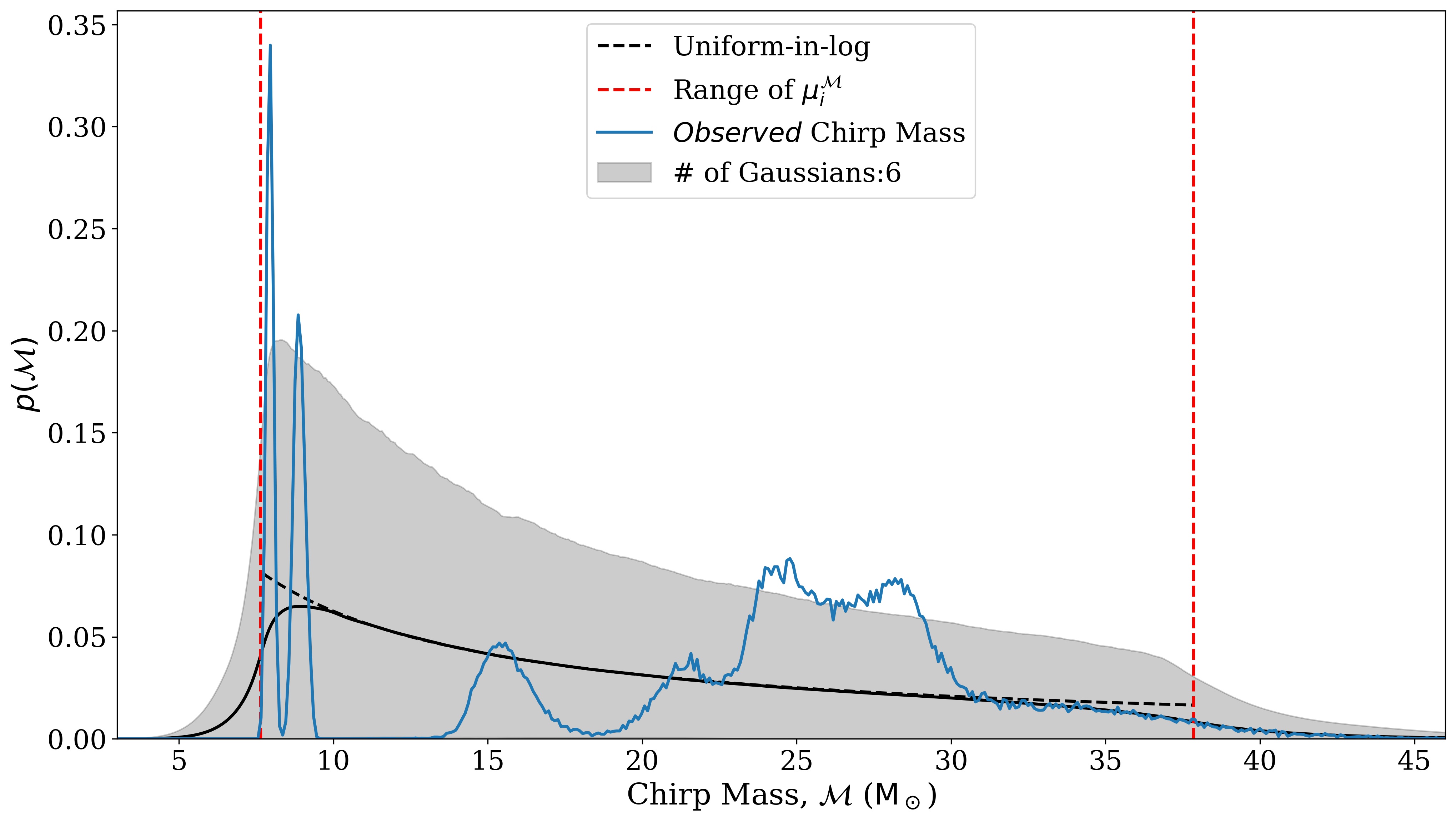}
\includegraphics[width=0.95\textwidth]{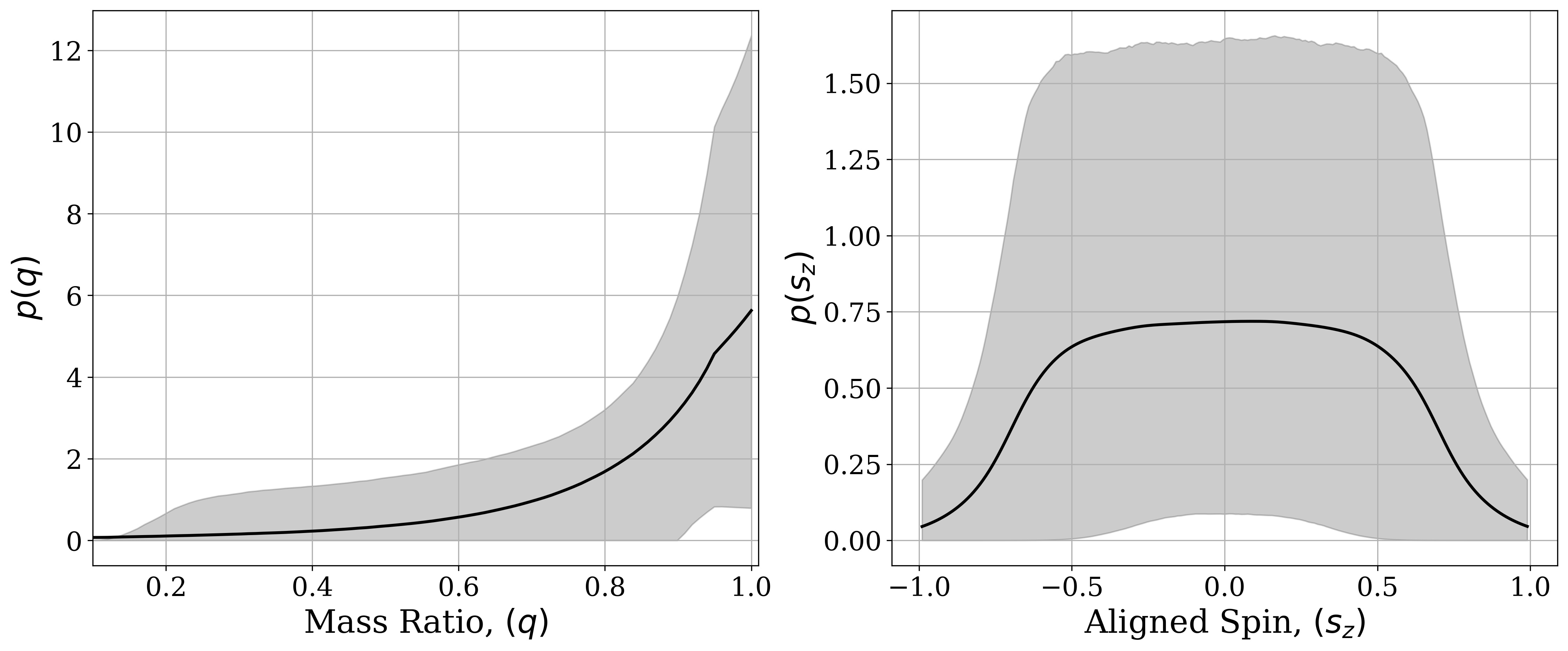}
\caption{Prior chirp mass, mass ratio, and aligned spin-distribution were obtained by performing the analysis with a flat likelihood with no smoothing constraints. Top) The shaded region is the 90\% credible interval for the chirp mass. The black curve is the mean distribution and the dashed black curve is the uniform-in-log distribution. The red dashed lines are the boundaries $\mathcal{M}_{s}^{\mathrm{min}}$ and $\mathcal{M}_{s}^{\mathrm{max}}$, and the blue curve is obtained by stacking the chirp mass estimates of all the observations into a histogram, bottom) The shaded 90\% credible interval for mass ratio and aligned-spin distribution corresponding to the priors used in this analysis. The curves are the mean distribution.}
\label{fig:priors}
\end{figure}

\begin{figure}
\centering
\includegraphics[width=0.95\linewidth]{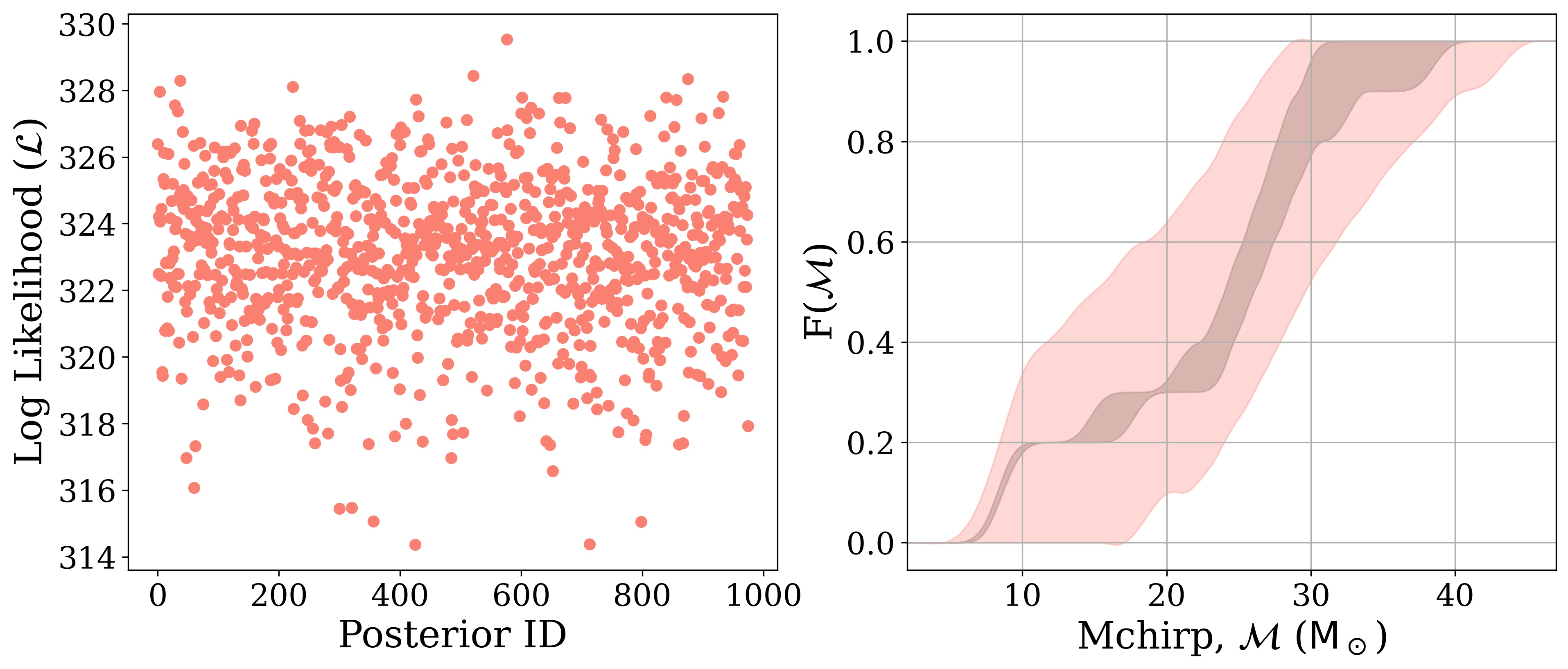}
\caption{Left) The natural log of the likelihood defined in equation \ref{eq:pop_bayes} showing no visible trend in their values indicate proper convergence of the sampler. Right) The salmon band is the 90\% confidence of the cumulative probability of the posterior predictive obtained after applying selection effects to the reconstructed chirp mass distribution. The grey band is the 90\% confidence obtained by bootstrapping various realisations of the observed data. Each realisation of the observed data is generated by re-weighting the chirp mass estimate of the observations to the reference population and selecting one data point from each one of them. The observed data is enclosed within the 90\% confidence of the posterior's prediction.}
\label{fig:sanity}
\end{figure}

Instead of listing the hyper-parameter ranges, we show the distribution of chirp mass, mass ratio, and aligned-spin corresponding to the priors used in Figure \ref{fig:priors}.
 We have performed the analysis using a number of components between 3 and 8. Table \ref{table:margL} lists the marginal log likelihood for these analyses. The marginal likelihood remains mostly unchanged for a wide range of component number\footnote{To calculate an approximate value for the marginal likelihood we use the prescription defined in \cite{doi:10.1198/016214501750332848}.}. We present results that use six components. Moreover, we have chosen to not apply any smoothing. We perform sanity checks to verify if the observed distribution is consistently predicted by the reconstructed distribution. We also check if the sampler has converged by observing the presence of any trend in the likelihood value of the posterior. Both of these checks are presented in figure \ref{fig:sanity}.
 
\begin{figure}
\centering
\includegraphics[width=0.95\textwidth]{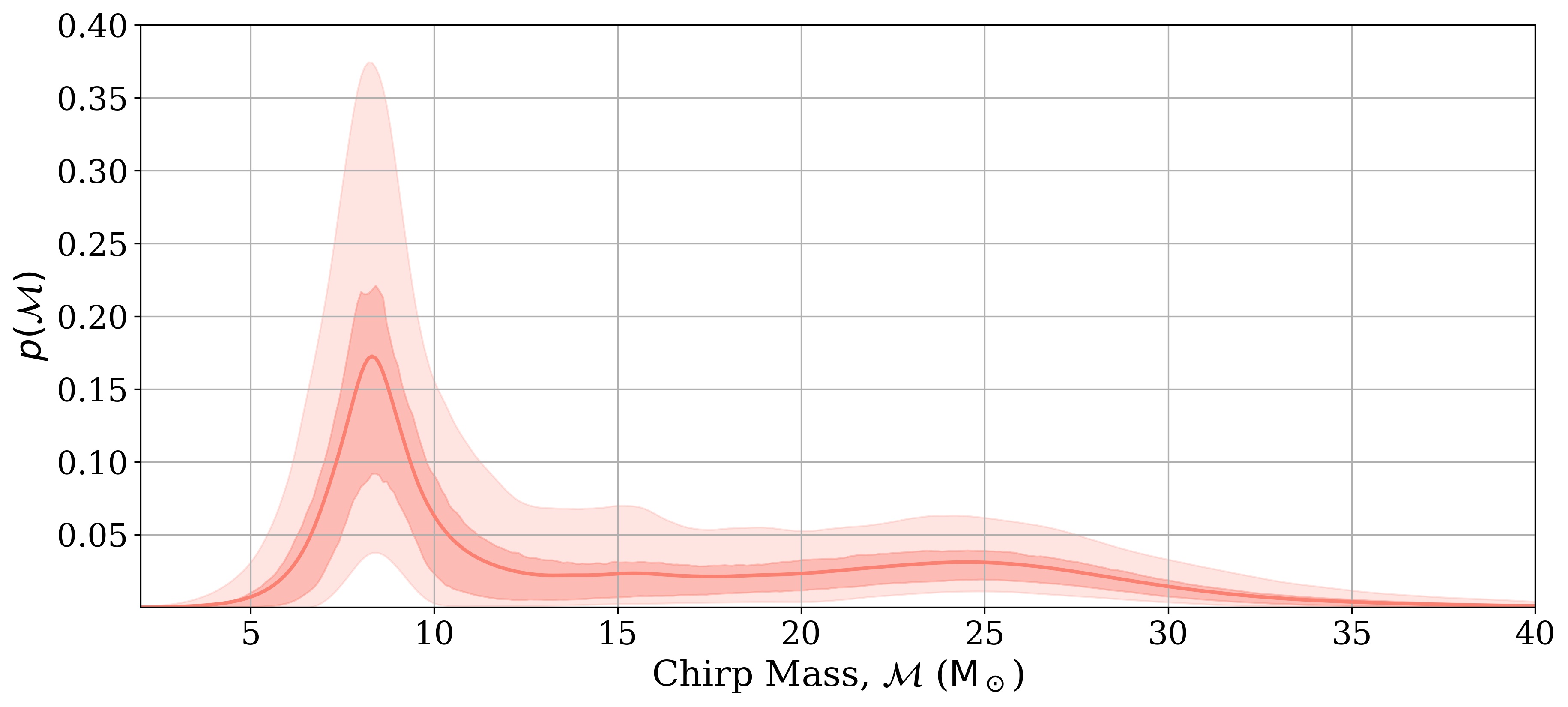}
\includegraphics[width=0.95\textwidth]{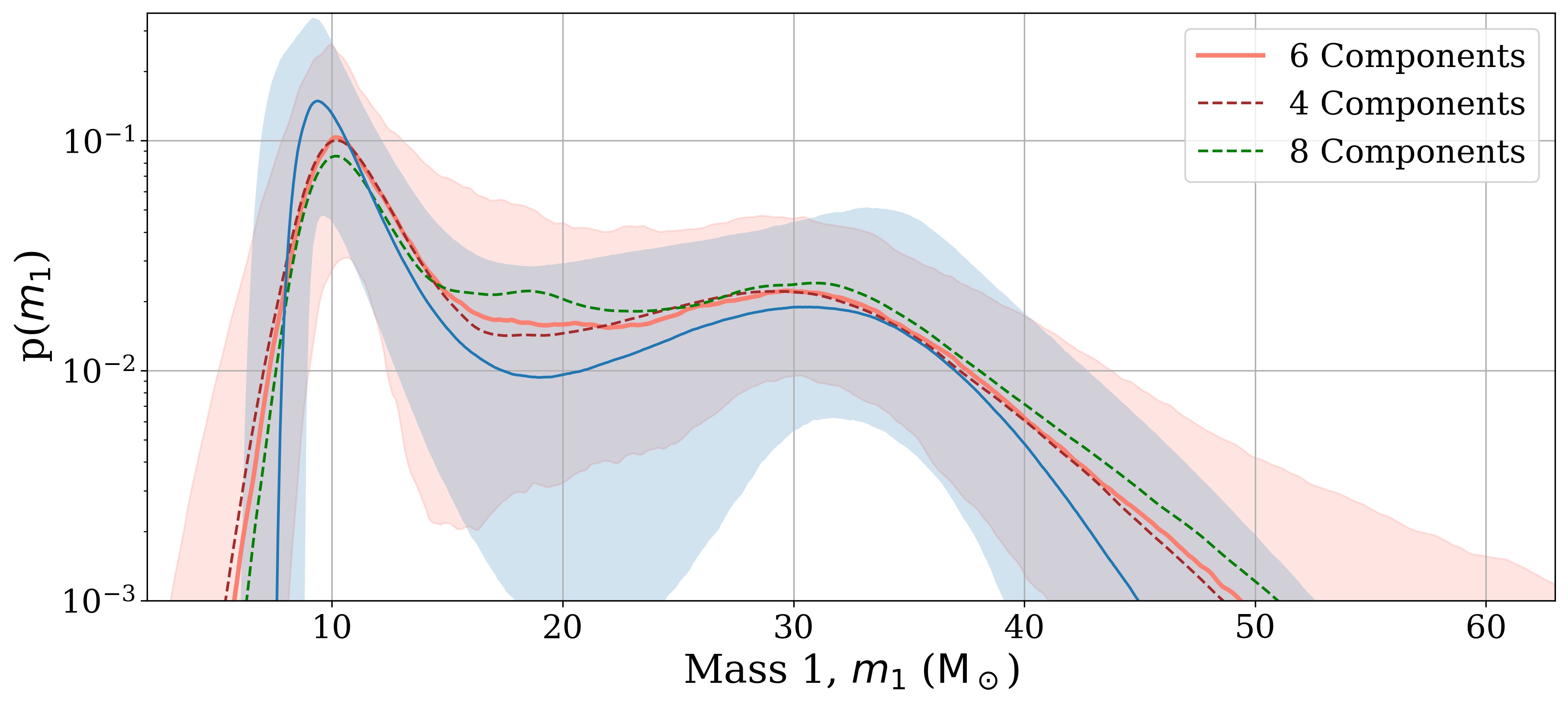}
\caption{Top) The reconstructed chirp mass distribution. The dark salmon band is the 50 \% credible interval, the light salmon band is the 90\% credible interval and the salmon curve is the mean distribution, bottom) The reconstructed primary mass distribution. The salmon band is the 90 \% credible interval and the salmon curve is the mean distribution. For comparison, the reconstructed primary mass corresponding to model C from the LIGO/Virgo analysis is shown in blue colour\cite{2019ApJ...882L..24A}. The figure also includes the mean distribution obtained by analysis using 4 and 8 components.}
\label{fig:post_masses}
\end{figure}

\begin{figure}
\centering
\includegraphics[width=0.95\textwidth]{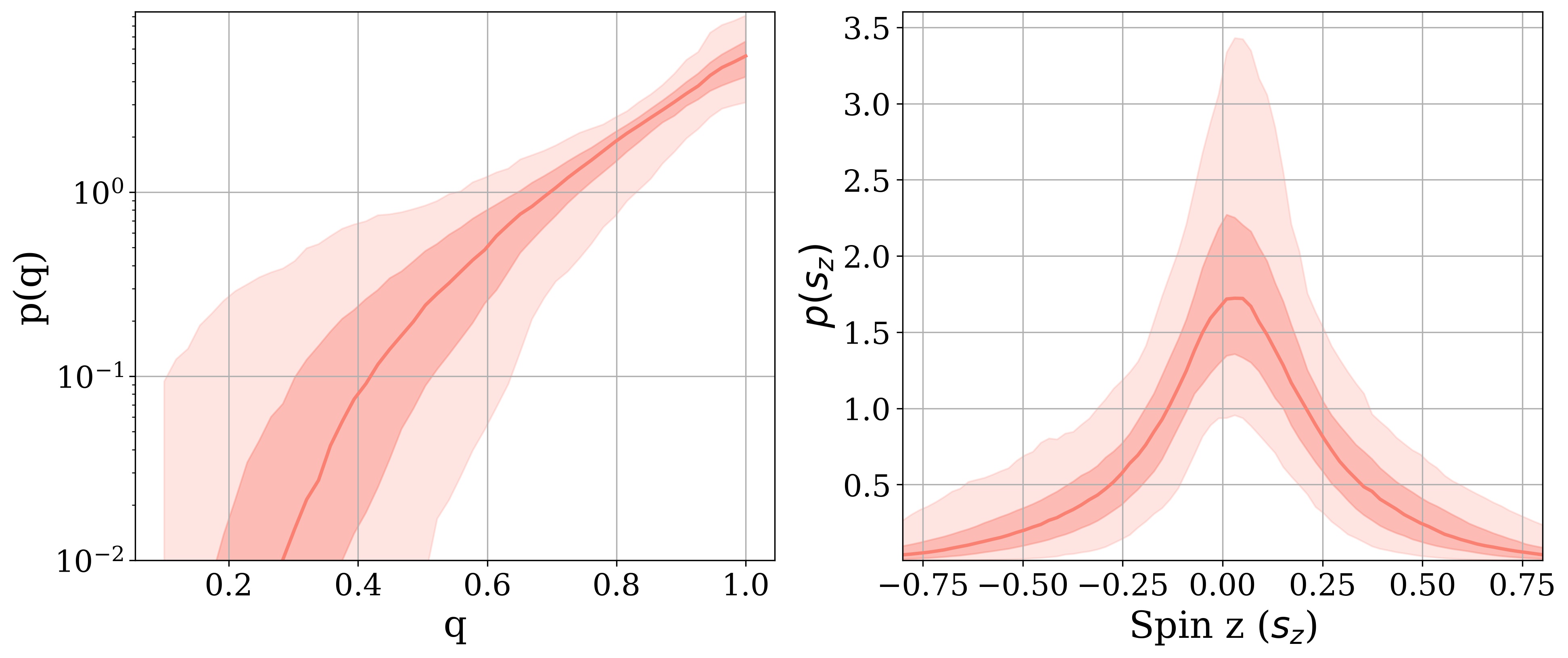}
\caption{The posterior on the mass ratio (q) and aligned spin component ($s_z$). The salmon band is the 90\% credible interval and the curve is the mean value. These distributions evidently favour closer to unity mass ratios and small magnitude for the spin-components aligned with the orbital angular momentum.}
\label{fig:post_q_sz}
\end{figure}

\begin{figure}
\centering
\includegraphics[width=0.95\textwidth]{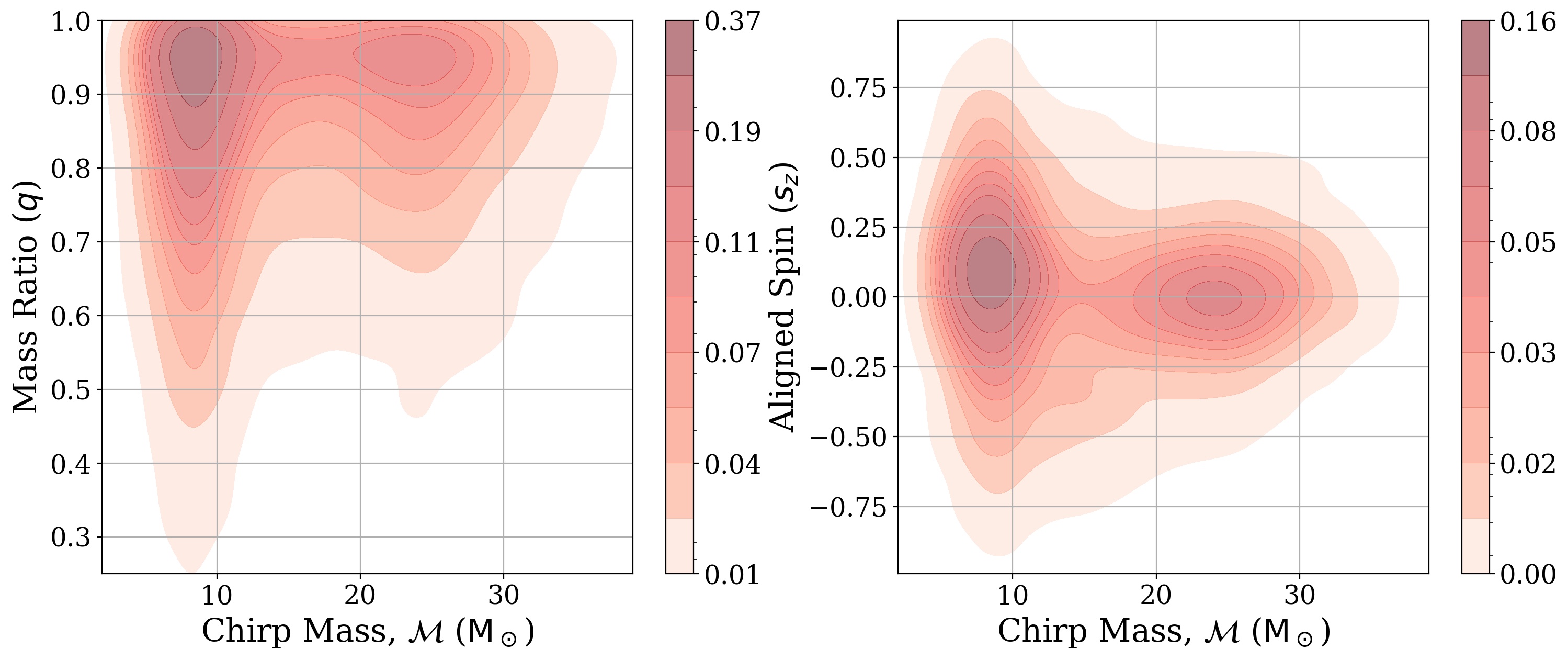}
\caption{The variation of the aligned-spin and mass ratio with the chirp mass of. Due to the higher spin of GW151226, there is support for a positive spin for low masses. For heavier masses the spins are small. The mass ratio remains close to one throughout the chirp mass range. There are 10 contours in each plot showing equally spaced confidence intervals with the first one for 5\% confidence and the last one for the 95\% confidence.}
\label{fig:mch_sz_q}
\end{figure}

\begin{figure}
\centering
\includegraphics[width=0.7\textwidth]{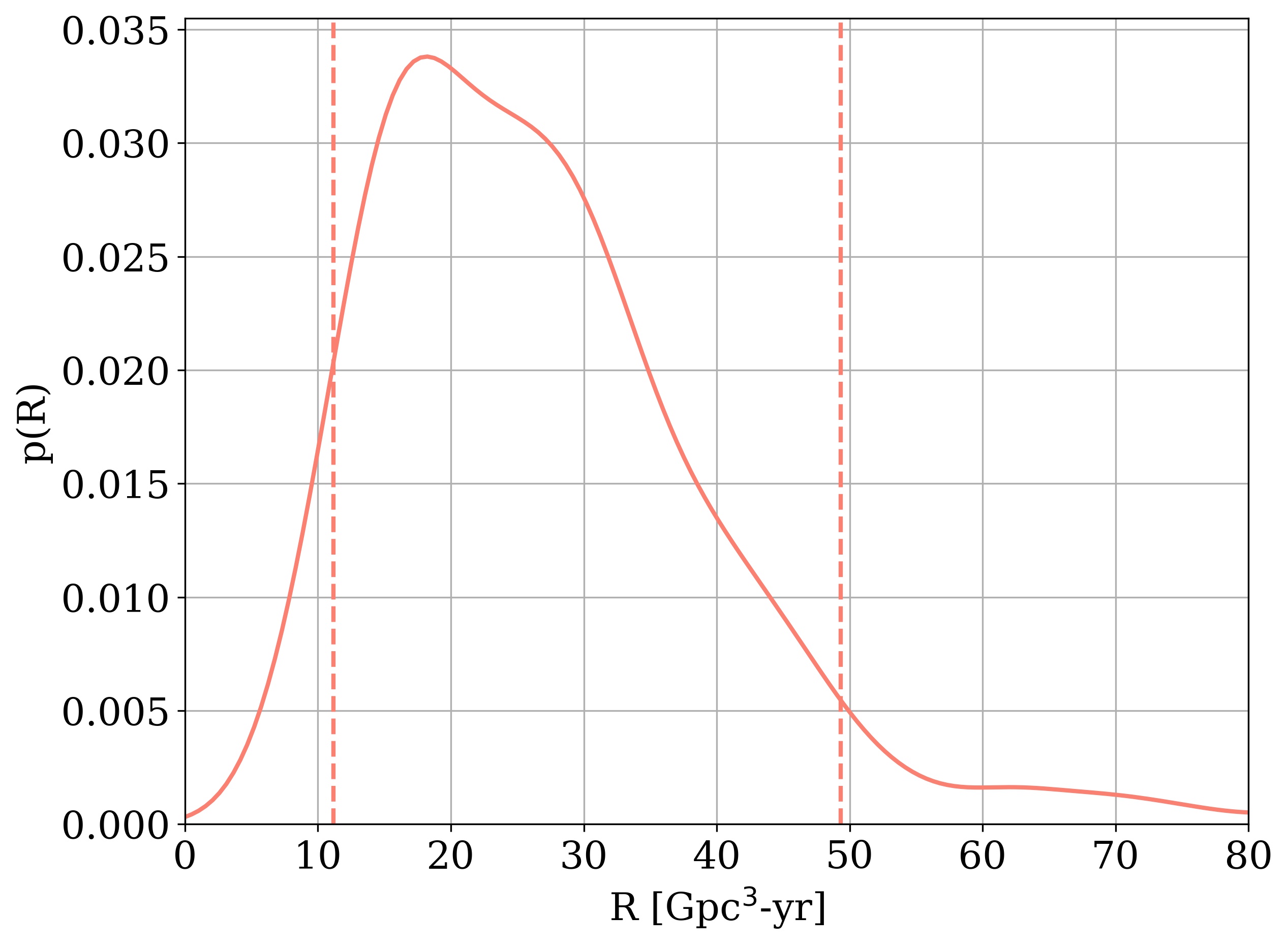}
\caption{The posterior on the measured rate. The 90\% confidence interval, enclosed within the dotted lines, is [11.2 - 49.3] $\mathrm{Gpc}^{-3}\mathrm{yr}^{-1}$.}
\label{fig:post_rate}
\end{figure}

 \begin{center}
\begin{table}
\begin{tabular}{ | m{4cm} | m{1.4cm} | m{1.4cm} | m{1.4cm} | m{1.4cm} | m{1.4cm} | m{1.4cm} |} 
\hline
\# of Components & 3 & 4 & 5 & 6 & 7 & 8\\ 
\hline
Marginal Likelihood & 324.3 & 324.5 & 324.5 & 324.5 & 324.4 & 324.2\\
\hline
\end{tabular}
\caption{Marginal Likelihood for analysis with different number of components. For a larger component number, marginal likelihood monotonically decreases. The marginal likelihood remains unchanged for a wide range of component number. We cannot conclude the most optimum component number and present results for reconstruction that uses six components.}
\label{table:margL}
\end{table}
\end{center}
Figure \ref{fig:post_masses} plots the chirp mass and the primary mass distribution. The primary mass distribution is obtained by making a variable transformation. The figure shows the reconstructed primary mass distribution obtained using "model C" in the LIGO/Virgo analysis \cite{2019ApJ...882L..24A}. It also shows effect of changing the number of components in the analysis. The reconstructed mean is consistent for most of the mass range except for the feature at around $18 M_\odot$ which is increasingly pronounced with the increase in the number  of components. This feature is primarily due to the observation GW151012. The 4 component analysis is most consistent with the LIGO/Virgo analysis but the 6 component analysis is most favoured. As the analysis is data driven it is expected that fluctuation in data can give rise to features in the reconstructed \emph{mean}. However, the significance of a feature can be estimated in various ways, e.g. i) by comparing the mean distribution with feature replaced by a best fit powerlaw, ii) by making a comparison with the mean reconstruction obtained using a phenomenological model, and iii) by comparing two reconstructions obtained using a different number of components with one that shows the presence of feature and other that does not. For the feature at $18 M_\odot$, the Bayes factor is only 1.2 between the 6 and 4 component analysis. Thus it is not noteworthy in the presented analysis.
Figure \ref{fig:post_q_sz} plots the mass ratio and aligned-component spin distributions. All the observations favour a mass ratio of closer to unity. Reconstruction suggests that the formation channels for black holes prefer producing equal mass binaries with the fractional contribution declining rapidly for lower mass ratios. The measured spins on all the observations are also small. The only exception being GW151226 which has a moderate spin magnitude. VAMANA facilitates the modeling of spins and mass ratio as dependent on the chirp mass. Figure \ref{fig:mch_sz_q} shows the variation of the aligned spin with the total mass of the binary. Except for GW151226, the spins are consistent with small magnitudes and do not vary with the chirp mass of the binary black holes. As has already been reported in multiple publications, this is in contrast to the black hole spins measured in x-ray binaries or the spins expected in the hierarchical merger scenario where black holes that acquired a remnant spin during their mergers go on and merge again with other black holes \cite{mm2015, Miller:2009cw, McClintock:2011zq, mcclintock-2006-652, gou2011}.
Finally figure \ref{fig:post_rate} plots the posterior on the merger rate, the 90\% confidence interval of which is $27.0^{+22.3}_{-15.8}\,\mathrm{Gpc}^{-3}\mathrm{yr}^{-1}$. 

\subsubsection{Effect of Smoothing}

\begin{figure}
\centering
\includegraphics[width=0.95\textwidth]{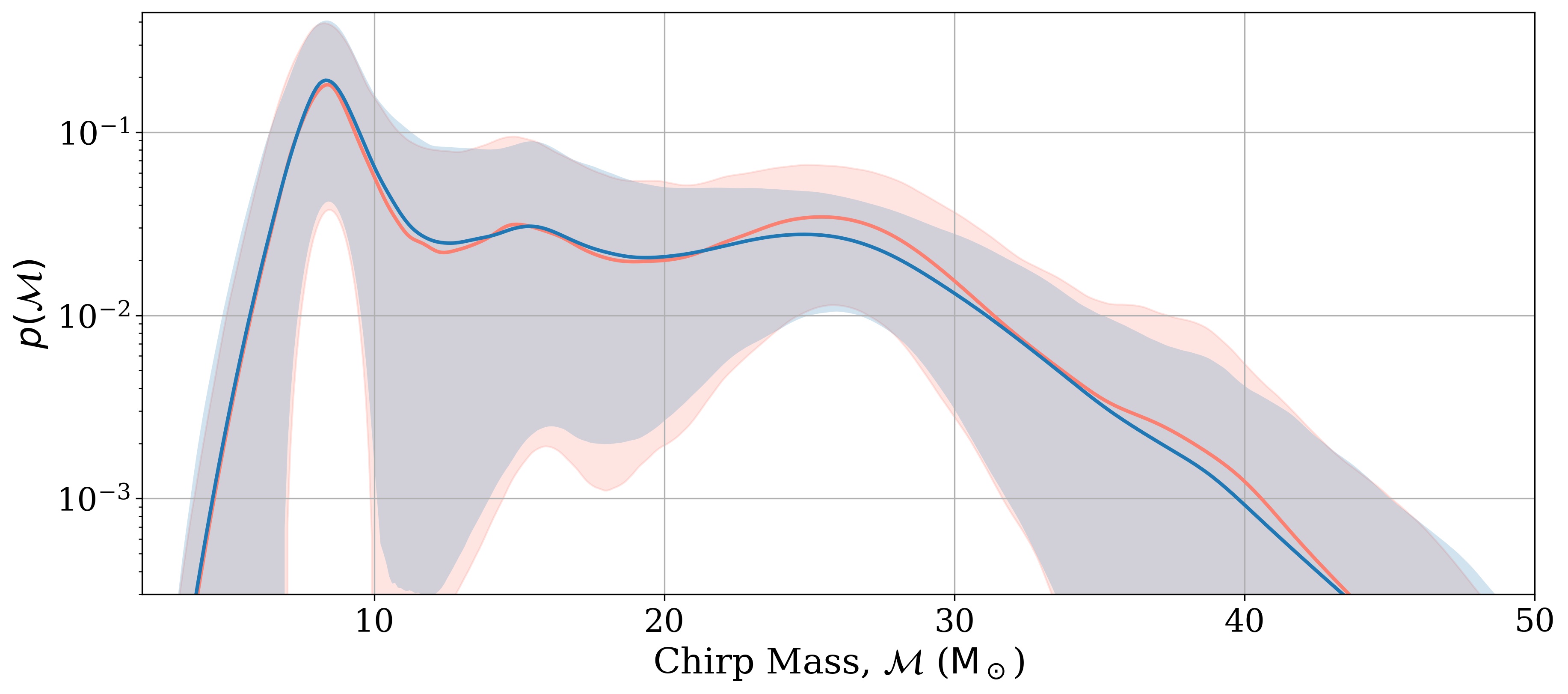}
\caption{The curves are the mean reconstructed chirp mass and the bands are the 90\% credible interval. The salmon plot has no smoothing applied, while the blue curve applies $r_{\mathrm{eff}} = 0.2$.}
\label{fig:smoothing}
\end{figure}

Figure \ref{fig:smoothing} shows the effect of smoothing on the reconstructed chirp mass distribution. The apparent effect is suppression of features that are not strongly supported by the data. An optimum smoothing threshold can be chosen by a bandwidth selection method operating under some rule-of-thumb. Alternatively, a value that maximises the marginal likelihood can also be used. For the analysis performed on the real data marginal likelihood is maximised for $r_{\mathrm{eff}} = 0.2$.

\section{Conclusion}
\label{conclusion}
In this article, we introduced VAMANA, a flexible scheme to model the properties of binary black hole population using a mixture model. We employ a mixture model in reconstructing the chirp mass, aligned-spin, and mass ratio distribution. We show that the analysis is capable of reconstructing complex distributions such as the power-law distribution and expect this flexible methodology will facilitate the extraction of any intricate features in the population. We did not introduce redshift as a signal parameter in this article but have proposed an extension in a separate article \cite{2020arXiv201208839T}. Moreover, this method can be extended to include binary neutron star and neutron star-black hole binaries but including low mass, compact binaries will further increase the dynamic range of the chirp mass distribution. A limited number of Gaussians will probably not be sufficient to model a density that changes by a few orders of magnitude over the chirp mass range. Alternatively, this analysis can be broken into two on the chirp mass range to model. We plan to include some of these developments in future works.

\section*{Acknowledgement}
The author would like to thank Stephen Fairhurst and Bangalore Sathyaprakash for many useful discussions. Much appreciation for the feedback provided by Joseph Mills that was quite helpful. This work was supported by the STFC grant ST/L000962/1.

We are grateful for the computational resources provided by Cardiff University, and funded by an STFC grant supporting UK Involvement in the Operation of Advanced LIGO.

The authors thank the LIGO Scientific Collaboration for access to the data and gratefully acknowledge the support of the United States National Science Foundation (NSF) for the construction and operation of the LIGO Laboratory and Advanced LIGO as well as the Science and Technology Facilities Council (STFC) of the United Kingdom, and the Max-Planck-Society (MPS) for support of the construction of Advanced LIGO. Additional support for Advanced LIGO was provided by the Australian Research Council.

\bibliography{apsbib}
\end{document}